# Electron transport and anisotropy of the upper critical magnetic field in a Ba$_{0.68}$K$_{0.32}$Fe$_2$As$_2$ single crystals

V.A. Gasparov[1], F. Wolff-Fabris[2], D.L. Sun[3], C.T. Lin[3], J. Wosnitza[2]

[1] Institute of Solid State Physics RAS, Chernogolovka, 142432, Russian Federation
[2] Hochfeld-Magnetlabor Dresden (HLD), Forschungszentrum Dresden-Rossendorf, 01314 Dresden, Germany
[3] Max Planck Institute for Solid State Research, 70569 Stuttgart, Germany

## Abstract

Early work on the iron-arsenide compounds supported the view, that a reduced dimensionality might be a necessary prerequisite for high–$T_c$ superconductivity. Later, however, it was found that the zero-temperature upper critical magnetic field, $H_{c2}(0)$, for the 122 iron pnictides is in fact rather isotropic. Here, we report measurements of the temperature dependence of the electrical resistivity, $\rho(T)$, in Ba$_{0.5}$K$_{0.5}$Fe$_2$As$_2$ and Ba$_{0.68}$K$_{0.32}$Fe$_2$As$_2$ single crystals in zero magnetic field and for Ba$_{0.68}$K$_{0.32}$Fe$_2$As$_2$ as well in static and pulsed magnetic fields up to 60 T. We find that the resistivity of both compounds in zero field is well described by an exponential term due to inter-sheet umklapp electron-phonon scattering between light electrons around the $M$ point to heavy hole sheets at the $\Gamma$ point in reciprocal space. From our data, we construct an $H$–$T$ phase diagram for the inter-plane ($H \parallel c$) and in-plane ($H \parallel ab$) directions for Ba$_{0.68}$K$_{0.32}$Fe$_2$As$_2$ compounds. Contrary to published data for underdoped 122 FeAs compounds, we find that $H_{c2}(T)$ is in fact anisotropic in optimally doped samples down to low temperatures. The anisotropy parameter, $\gamma = H^{ab}_{c2}/H^{c}_{c2}$, is about 2.2 at $T_c$. For both field orientations we find a concave curvature of the $H_{c2}$ lines with decreasing anisotropy and saturation towards lower temperature. Taking into account Pauli spin paramagnetism we perfectly can describe $H_{c2}(T)$ and its anisotropy.

## Introduction

A new "iron" era in the research of high-temperature superconductivity (HTSC) was initiated by Kamihara *et al*. [1], when reporting a superconducting transition at $T_c \approx 4$ K in the layered iron-based oxy-pnictide LaOFeP. Soon after, a series of LaOMPn superconducting compounds were synthesized (with M = Mn, Fe, Co, or Ni and Pn a pnictogen: P, As, or Sb), which led to a new boom in HTSC [2]. Although the Fermi surfaces of the iron pnictides show a strongly anisotropic character [2], reports on the anisotropy of the upper critical field, $H_{c2}(T)$, are quite contradictory [2-10]. Investigations in relatively weak fields (up to 10 T) have revealed a significant anisotropy of $\gamma = H^{ab}_{c2}(0)/H^{c}_{c2}(0) \approx 4$ in NdFeAsO$_{0.9}$F$_{0.1}$[6] and $\gamma$ from 1.8 in NaFeAs with Co and P doping [7] up to 3.4 in Ba(Fe$_{0.92}$Co$_{0.08}$)$_2$As$_2$ [8]. In these fields, $H_{c2}(0)$ is evaluated by the slope of d$H_{c2}$/d$T|_{Tc}$ close to $T_c$ according to the well-known Werthamer – Helfand –Hohenberg (WHH) model for the orbital critical magnetic field $H^{orb}_{c2}(0) \approx -0.69 T_c$ (d$H_{c2}$/d$T)|_{Tc}$ [11]. However, direct measurements of $H_{c2}$(T) in pulsed magnetic fields have shown that the actual anisotropy of $H_{c2}(0)$ becomes very small at low temperatures [3,4].

Another striking property of the Fe pnictides is the temperature dependence of the resistivity, $\rho(T)$. A characteristic feature of $\rho(T)$ in 122 hole-doped Ba(Eu)$_x$K$_{1-x}$Fe$_2$As$_2$ single crystals is the close-to-linear dependence of $\rho(T)$ with a tendency to saturation at high temperatures. Actually, there is no consensus about the nature of this behavior [2]. Such a linear dependence is characteristic for a magnetic quantum phase transition [12]. A linear $\rho(T)$ dependence with no saturation was observed in the electron-doped iron pnictides $SmO_xF_{1-x}FeAs$ [13] and $Ba(Fe_{1-x}Co_x)_2As_2$ [14]. As was shown by Rullier-Albenque *et al.*, the extraction of the pure electron conductivity in $Ba(Fe_{1-x}Co_x)_2As_2$ and $Ba(Fe_{1-x}Ru_x)_2As_2$ is strongly complicated due to the complex temperature dependences of the electron $n_e(T)$ and hole $n_h(T)$ number of carriers [15]. Even though some information on these quantities can be gained from Hall-effect studies, the origin of the temperature dependences of $\rho(T)$ remains unclear.

Here, we show that for $Ba_xK_{1-x}Fe_2As_2$ single crystals the whole $\rho(T)$ dependence can be explained by the freezing out of electron-phonon inter-sheet umklapp scattering between light electron pockets and heavy hole cylinders of the Fermi surface. We present electrical-resistivity measurements in high-quality $Ba_{0.5}K_{0.5}Fe_2As_2$ and $Ba_{0.68}K_{0.32}Fe_2As_2$ single crystals for zero field and for $Ba_{0.68}K_{0.32}Fe_2As_2$ as well in magnetic fields up to 60 T. The latter data show an anisotropy of $H_{c2}(T)$, different from previously published data.

## Experimental

Single crystals of $Ba_{0.5}K_{0.5}Fe_2As_2$ and $Ba_{0.68}K_{0.32}Fe_2As_2$ were grown from FeAs flux in a zirconia crucible sealed in a quartz ampoule under argon atmosphere. A mixture of Ba, K, Fe, and As in an appropriate weight ratio was heated in a box furnace up to a maximum temperature of 980°C applied in order to soak the pure compounds. A cooling rate of 3°C/h was then applied to decrease the temperature to 550°C. The grown crystals were then decanted from the flux. The chemical composition was determined by energy-dispersive X-ray spectrometry (EDX) as well as inductively coupled plasma spectroscopy (ICP). The crystal morphology was studied by scanning electron microscopy (SEM). More details on the growth method, the crystal structure, and the characterization have been described elsewhere [8,16]. Bulk superconductivity was confirmed by magnetic-susceptibility and dc-conductivity measurements.

The sample resistances were measured using a four-probe van der Pauw technique from room temperature down to 4.2 K. The samples were plates with dimensions of about 1x1x0.15 $mm^3$ ($Ba_{0.68}K_{0.32}Fe_2As_2$) and 2x1x0.2 $mm^3$ ($Ba_{0.5}K_{0.5}Fe_2As_2$). X-ray diffraction revealed that the crystal surfaces are normal to the $c$ axis. The contacts at each sample corner were prepared with conducting silver paste and Au wires. The contact resistance was ~ 1 $\Omega$. The in-plane DC resistivity, $\rho_{ab}$, was measured by use of a Quantum Design Physical Property Measurement System (PPMS) at several fixed temperatures for magnetic fields below 14 T performing slow cooling and field-sweep rates. The current used for the PPMS measurements was 1 mA.

The magnetic-field dependence of $\rho_{ab}$ was measured at fixed temperatures using a 60 T capacitor-bank-driven pulsed magnet at the Hochfeld-Magnetlabor Dresden, with pulse durations of 150 ms. The current and frequency used throughout the pulsed-field measurements were 10 mA and 70 kHz, respectively. Also, while the data presented here were taken at the decaying part of the pulse, they were reproduced at the rising part, which confirms that our data are not affected by sample heating during the pulse. Nevertheless, the data shown here are from the down sweep of the field as the decay time of the pulsed field is much longer than its rise time. During the measurements, magnetic fields were applied along the $c$ axis and in the $ab$ plane while maintaining the current flow normal to the magnetic field.

## Results and discussion

Figure 1 shows the temperature dependence of the resistivity of both samples. For $Ba_{0.68}K_{0.32}Fe_2As_2$, $T_c = 38.5$ K is found with a very narrow transition width of $\Delta T_c = 0.3$ K, determined from the 10% and 90% resistance drop at $T_c$. Very sharp superconducting transitions were confirmed by magnetic-susceptibility measurements. Shielding fractions close to 1 demonstrate the bulk nature of superconductivity. The residual resistivity ratio RRR = R(300 K)/R(T$_{c,onset}$) of 16 for $Ba_{0.68}K_{0.32}Fe_2As_2$ is four times larger than the value of 4.3 for $Ba_{0.5}K_{0.5}Fe_2As_2$. This indicates a strongly reduced impurity scattering in our $Ba_{0.68}K_{0.32}Fe_2As_2$ sample compared to $Ba_{0.5}K_{0.5}Fe_2As_2$. Indeed, from wavelength dispersive X-ray spectroscopy on a cleaved $Ba_{0.5}K_{0.5}Fe_2As_2$ sample we found a depth-dependent Ba/K ratio from the nominal 0.5/0.5 on the surface to 0.7/0.3 in the bulk [2]. The $Ba_{0.68}K_{0.32}Fe_2As_2$ single crystals have the lowest residual specific heat reported so far for FeAs-based superconductors, indicating the superior quality and high purity of our samples. The ratio of the residual electronic specific heat to the normal-state value yields an estimate of the non-superconducting fraction of less than 2.4 % [17].

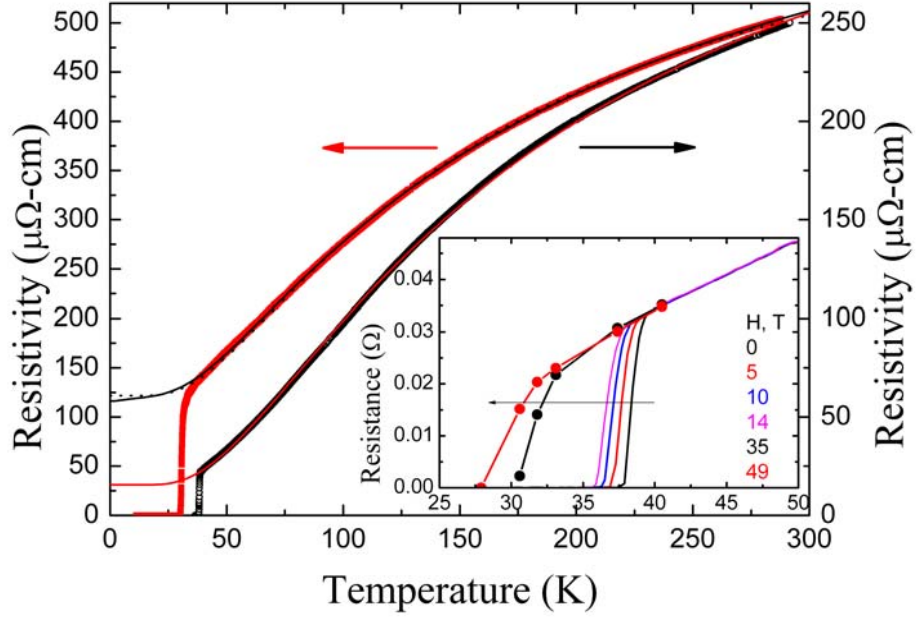

Fig.1. (Color online) Temperature dependence of the in-plane resistivity, $\rho_{ab}(T)$, of $Ba_{0.5}K_{0.5}Fe_2As_2$ (left axis, upper curve) and $Ba_{0.68}K_{0.32}Fe_2As_2$ (right axis, lower curve) single crystals in zero magnetic field. The solid curves show the results of best fits using Eq. (1) with $\rho(0) = 15.5$ μΩcm, $\rho_1 = 0$, $C = 406$ μΩcm, $T_0 = 158$ K for $Ba_{0.68}K_{0.32}Fe_2As_2$ (red line); $\rho(0) = 115.3$ μΩcm, $\rho_1 = 0.49$ μΩcm, $n = 0.83$, $C = 531$ μΩcm, $T_0 = 134$ K (black line) and $\rho_1 = 0$ (dotted line) for $Ba_{0.5}K_{0.5}Fe_2As_2$. The inset shows the temperature dependence of the resistance for a cleaved piece of the $Ba_{0.68}K_{0.32}Fe_2As_2$ single crystal at different magnetic fields parallel to the $ab$ plane. The data up to 14 T are from PPMS measurements, while the points are extracted from pulsed-field measurements.

A characteristic feature of the data shown in Fig. 1 is the quasi-linear temperature dependence of $\rho(T)$ above $T_c$ with a tendency to saturation towards high temperatures. The range where this quasi-linear dependence holds reduces for K concentrations below a critical value of $x_c = 0.4$ [2]. Various assumptions have been made regarding the saturation of $\rho(T)$ related both to the Ioffe–Regel limit, when the mean free path $l$ is comparable to the inter-atomic distance, and to the two-band conductivity [18]. However, as it was observed from electrical-transport measurements at high pressure, the position where the resistivity saturates shifts to lower temperatures as pressure increases, i.e. with decreasing the lattice parameter (see references in [2]). This observation contradicts the assumption of the Ioffe–Regel limit. Furthermore, we did not succeed to fit the $\rho(T)$ dependence by the sum of two different Bloch–Grüneisen conductivities with different Debye temperatures.

Thus, we have analyzed the resistivity data in terms of various theories proposed to describe the temperature dependence of the resistivity in superconducting A15 compounds [19-21]. It was suggested by Milewits *et al.* [21] that instead of the Bloch–Grüneisen or Wilson formula [19], $\rho(T)$ in those compounds may be described by the simple empirical equation:

$$\rho(T) = \rho_0 + \rho_1 T^n + C\exp(-T_0/T), \qquad (1)$$

where the third term arises from phonon-assisted scattering between two Fermi-surface sheets. Here, $\rho_0$ is the residual resistivity due to impurities and lattice defects, and $\rho_1$, $C$, and $T_0$ are constants. There are different possible origins for the second term discussed [20,21], such as electron-electron scattering ($n = 2$), $s$-$d$ scattering in the high-temperature limit of the Wilson model ($n = 1$) [19], or quantum critical behavior [12].

In Fig. 1, we show by solid lines the best fits over the full temperature range of $\rho(T)$ using Eq. (1) with $\rho_1 = 0$ for $Ba_{0.68}K_{0.32}Fe_2As_2$ and $\rho_1 = 0.49~\mu\Omega cm$ and $n = 0.83$ for $Ba_{0.5}K_{0.5}Fe_2As_2$. The dotted line shows a fit with $\rho_1 = 0$. It is evident that both fits are of almost equal quality, i.e., the second term in Eq. (1) is not relevant in our case. Most important is the last term, that explains the saturation of $\rho(T)$ at high temperatures (Fig. 1). Remarkably, Eq. (1), not previously discussed for the Fe pnictides, fits the data very well over the full temperature scale.

As was suggested by Wilson [19], the exponential term in Eq. (1) arises from inter-band electron-phonon umklapp scattering from a low-mass $s$-band sheet to a heavy-mass $d$-band sheet. The freezing out of these processes at low temperatures is responsible for the exponential dependence of $\rho(T)$. Such a scenario for electron-phonon umklapp scattering between a small and large Fermi-surface sheet was considered by one of us (V.A.G.) in order to determine the electron-phonon scattering rate [22,23]. In this model, $T_0 = \hbar\Delta k s/k_B$ is the temperature at which the averaged phonon wave-vector $q$ is equal to the inter-sheet distance $\Delta k$. Thereby, $s$ is the sound velocity and $k_B$ the Boltzmann constant.

We determined the sound velocity $s$ from inelastic x-ray-scattering data and calculated phonon frequencies for $BaFe_2As_2$ (no significant doping dependence has been found in the experiments) [24]. We obtained $s_t = 3.5 \times 10^5$ cm/sec for a transversal phonon branch and $s_l = = 5.9 \times 10^5$ cm/sec for a longitudinal. Keeping in mind that, the resistivity is due to electron scattering on longitudinal phonons [23], we will consider only $s_l$ here. Angular-resolved photoemission spectroscopy (ARPES) data [25-27] of hole-doped $Ba_{1-x}K_xFe_2As_2$ reveal four Fermi-surface sheets, two concentric corrugated cylindrical hole barrels around the $\Gamma$ point (with Fermi wave vectors $k_{F\alpha} \approx 0.2\pi/a$ for the $\alpha$ sheet and $k_{F\beta} \approx 0.45\pi/a$ for the $\beta$ sheet) and two electron tubes around the M points (with almost identical $k_F \approx 0.2\pi/a$ for the $\gamma$ and $\delta$ sheets). The hole sheets expand with increasing K concentration from the optimally doped $x = 0.4$ to $x = 0.7$, while the two electron sheets shrink a little bit with $x$ [25] and, thus, the inter-sheet hole-electron distance $\Delta k$ decreases. The estimates for the effective masses of these bands obtained from ARPES data are $m^*_\alpha = 4.8$, $m^*_\beta = 9.0$, $m^*_\gamma = 1.3$, and $m^*_\delta = 1.3$ in units of the free-electron mass $m_0$ [26]. Thus, in the Wilson two-band $s$-$d$ model the hole $\alpha$ and $\beta$ sheets with low mobility can be regarded as high-mass $d$ bands, while the electron $\gamma$ and $\delta$ sheets are equivalent to low-mass $s$ bands.

Using the longitudinal sound velocity mentioned above and the inter-sheet distance $\Delta k = 0.35\pi/a$ from [26], we obtain $T_0 = 127$ K. These estimates not only reasonably agree with the experimental data but also quantitatively follow the scenario of the $\Delta k$ change with doping. Thus, we have shown that the temperature variation of $\rho(T)$ in Fe pnictides can be fit by a surprisingly simple expression without invoking non-Fermi-liquid speculations.

Now, we turn to the temperature dependence of the upper critical field of $Ba_{0.68}K_{0.32}Fe_2As_2$. $H_{c2}(T)$ was determined from the resistive transitions using as a criterion the 50% resistance drop from the normal-state resistance. Figure 2 shows the magnetic-field dependence of the resistance in pulsed magnetic fields up to 60 T. The data show a rather small magnetoresistance and a very high $H_{c2}$, corresponding to the high $T_c = 38.5$ K. The superconducting transitions are not substantially broadened, but just move to higher fields with decreasing temperature. The temperature dependence of the resistance, extracted from these data at two magnetic fields, is shown in the inset of Fig. 1. The data in pulsed fields nicely agree with data taken in static field. The normal-state resistance above $T_c$ displays no field-orientation dependence. With decreasing temperature the high-field resistance decreases monotonously for both in-plane and out-of-plane fields.

Figure 3 shows the temperature dependences of $H_{c2}{}^{ab}$ and $H_{c2}{}^{c}$ in almost optimally doped $Ba_{0.68}K_{0.32}Fe_2As_2$ ($T_c$ = 38.5 K) with the magnetic field aligned within ($H \parallel ab$) and perpendicular ($H \parallel c$) to the basal plane, respectively. Data taken by use of the PPMS as well as by pulsed-field measurements are shown. As is evident in Fig. 3, the conventional linear behavior of $H_{c2}$ vs $T$ is observed close to $T_c$, with clearly different slopes for the two field orientations. Towards lower temperatures, a clear saturation for both field orientations is evident. The anisotropy parameter $\gamma = H_{c2}{}^{ab}/H_{c2}{}^{c}$ is about 2.2 near $T_c$, but decreases at low temperatures considerably. These results disagree somewhat with $H_{c2}$ data for underdoped $Ba_{0.55}K_{0.45}Fe_2As_2$ ($T_c$ = 32 K) [28] and $Ba_{1-x}K_xFe_2As_2$ with $x \approx 0.4$ ($T_c$ = 28 K) [3]. In these studies a convex shape of $H_{c2}(T)$ for $H \parallel ab$ was observed, qualitatively in line with our results. Contrary to our data, however, $H_{c2}(T)$ for $H \parallel c$ follows an almost linear temperature dependence down to 14 K [28] and 10 K [3]. Surprisingly small anisotropies, $\gamma$, were observed. This was claimed to be a consequence of the three-dimensional Fermi-surface topology, although ARPES studies clearly show a highly anisotropic character of the Fermi surface in these pnictides [2,25-27]. The anisotropy we observe close to $T_c$ is in good agreement with the data reported by Welp *et al*. for $Ba_{0.6}K_{0.4}Fe_2As_2$ ($T_c$ = 34.6 K) [6] and Sun *et al*. for $Ba_{0.68}K_{0.32}Fe_2As_2$ [8] at low fields. At the same time, the anisotropies of $H_{c2}(0)$ in oxifluoride Fe 1111 compounds are substantially higher than for the 122 materials, and a transition to the normal state at low temperatures is not achieved even in pulsed fields above 60 T [2,6].

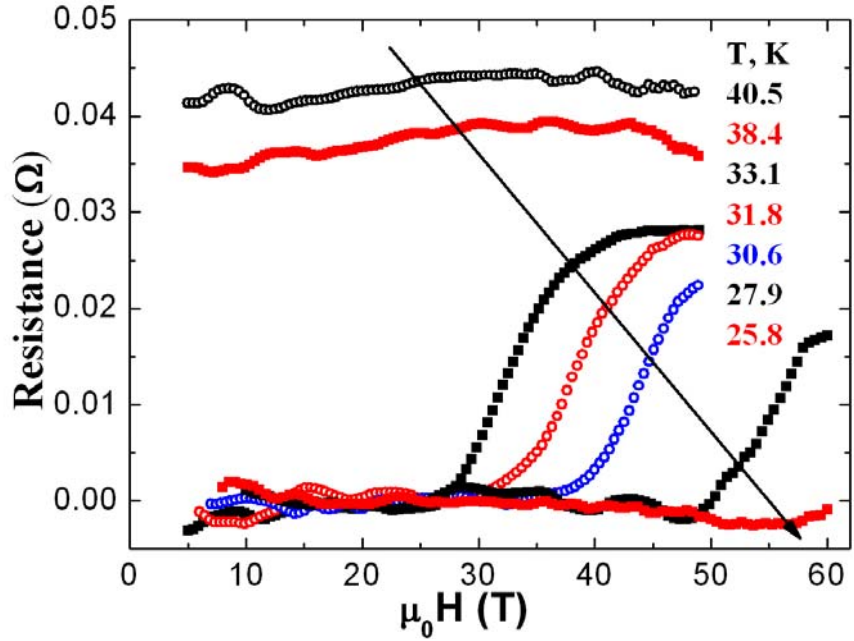

Fig. 2. (Color online) In-plane resistance ($\rho_{ab}$) of $Ba_{0.68}K_{0.32}Fe_2As_2$ vs pulsed magnetic fields for various temperatures. The magnetic field was applied parallel to the $ab$ plane.

The temperature-dependent anisotropy $\gamma$ we observed is most likely due to two independent superconductivity destruction mechanisms [2,10]: (i) at higher temperatures, Cooper pairing is suppressed by orbital currents that screen the external field; (ii) towards lower temperatures, the limiting effect is caused by the Zeeman splitting, i.e., when the Zeeman energy becomes larger than the condensation energy the Pauli limit, $H_p$, is reached [29,30]. In the most simple approximation $H_p$ is given by 1.84 [T/K] $T_c$ [29], giving $H_P$ = 71 T. In our case, this paramagnetic limit is lower than the orbital limit $H^*_{c2}(T)$ which is related to the slope of $H_{c2}(T)$ close to $T_c$. With the experimental slopes $dH^c_{c2}/dT = -4.55$ T/K and $dH^{ab}_{c2}/dT = -9.97$ T/K for $H \parallel c$ and $H \parallel ab$, respectively, the WHH model [11] predicts orbital-limited fields of $H^{*c}_{c2}(0) =$ 121 T and $H^{*ab}_{c2}(0) = 265$ T at $T = 0$. The dotted lines in Fig. 3 show the orbital critical fields within the WHH approach for both field orientations ignoring the Pauli limiting.

When including Pauli paramagnetism, the upper critical field is reduced relative to $H^*_{c2}(T)$ to:

$$H_{c2}(T) = H^*_{c2}(T)/[1 + \alpha^2(T)]^{1/2}, \qquad (2)$$

with the Maki parameter $\alpha(T) = \sqrt{2}\ H^*_{c2}/H_P$. Indeed, when using this equation with the orbital critical fields estimated above, we can perfectly describe the experimental data for both field orientations with only one free parameter, namely $H_p = 95.7$ T (solid lines in Fig. 3). This value is somewhat larger than the simple estimate of about 71 T. This, however, is not unexpected since in the latter value no many-body correlations and strong-coupling effects are included. A better estimate for $H_P$ could be extracted e.g. from the superconducting energy gap at zero temperature [29]. The latter value, however, is not known for $Ba_{0.68}K_{0.32}Fe_2As_2$. Anyway, the Pauli limit of 97.5 T extracted from our data is very reasonable and leads to the excellent one-parameter fit shown in Fig. 3.

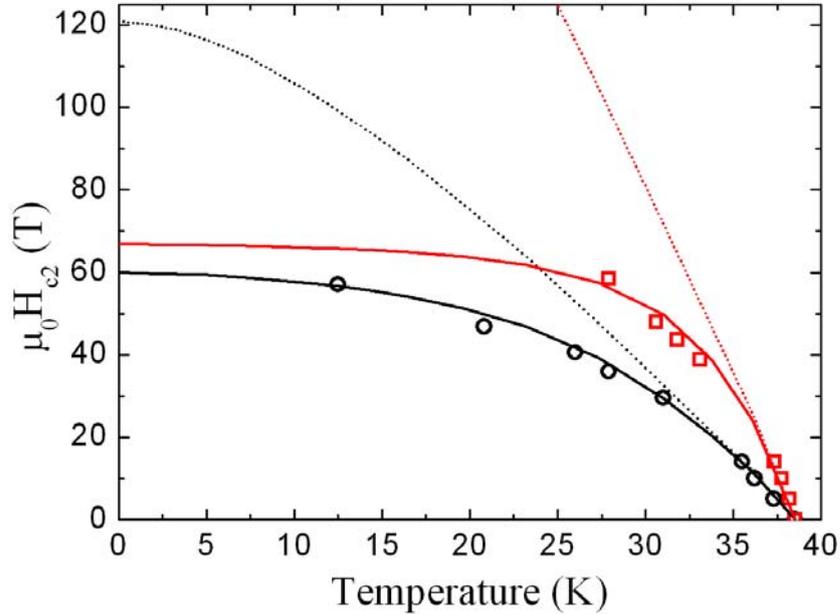

Fig.3. (Color online) Temperature dependences of the upper critical fields, $H_{c2}(T)$, of $Ba_{0.68}K_{0.32}Fe_2As_2$ for $H \parallel c$ (black circles) and $H \parallel ab$ (red squares). The dotted lines indicate the temperature dependences according to the WHH model neglecting Pauli-limiting ($\alpha = 0$). The solid lines show the dependences including Pauli pair breaking using Eq. (2) with only one free parameter, $H_p = 97.5$ T, for both field orientations, and with the Maki parameters $\alpha_c(0)=1.75$ (black line) and $\alpha_{ab}(0)=3.85$ (red line).

*In conclusion*, we have shown that the temperature dependence of the electrical resistivity in 122 hole-doped Fe pnictides can be very well described by the surprisingly simple Eq. (1). Thereby, an exponential term takes into account the freezing out of inter-sheet electron-phonon scattering between light-electron and heavy-hole Fermi-surface sheets, without the necessity to invoke non-Fermi-liquid scenarios. We further determined the upper critical field in a nearly optimally doped $Ba_{0.68}K_{0.32}Fe_2As_2$ single crystal up to 60 T. While the orbital-limited upper critical fields show an appreciable anisotropy, Pauli spin paramagnetism substantially limits $H_{c2}(T)$ at lower temperatures and, correspondingly, also the anisotropy. We consistently could describe our $H_{c2}(T)$ data with one fitting parameter, i.e., the Pauli-limiting field, over the whole temperature region for both field orientations.

**Acknowledgment**

We would like to thank V.F. Gantmakher, R. Huguenin, and A.B. Boris for helpful discussions and S.S. Khasanov for the elemental analysis of the samples by wavelength dispersive x-ray spectroscopy. Part of this work has been supported by EuroMag NET II under the EU contract 228043, the German Science Foundation within SPP 1458, and the RAS Program: New Materials and Structures (Grant 4.13).